\documentclass[
  reprint,
  amsmath,amssymb,
  aps,
  pre,
  floatfix,
  longbibliography
]{revtex4-2}

\usepackage{graphicx}
\usepackage{dcolumn}
\usepackage{bm}
\usepackage{hyperref}

\begin{document}

\title{Dissipative Avalanche Regimes Driven by\\Memory-Biased Random Walks on Networks}

\author{Mohammad Jafari}
\email{mojafari@id.uff.br}
\affiliation{%
  Universidade Federal Fluminense,\\
  Niter\'{o}i, Brazil
}

\begin{abstract}
We investigate a network model in which a single random walker combines local diffusion
with preferential resetting to previously visited nodes.  Each arrival deposits one unit
of stress on the target node, and threshold crossings trigger sandpile-like relaxation
cascades.  The fixed per-neighbor transfer rule produces a brittle transition on
Watts--Strogatz networks: below the stress-balance condition $\alpha k \simeq T$ cascades
remain short, whereas mildly supercritical transfer values generate runaway-capped events
at large system sizes.  A subtractive dissipative rule---in which a toppling node loses
$T$ units and redistributes only $\beta T$ across its neighbors---stabilizes broad,
finite cascades over a significantly wider parameter range.  For $\beta = 0.995$ and
$0.998$, the dissipative model remains non-runaway through $N = 4096$ and favors
power-law tails by AIC model selection; however, system-scale event fractions decrease
with $N$, a branching-ratio proxy remains below unity, and bootstrap Kolmogorov--Smirnov
tests reject a pure power law.  Shuffled-order controls that preserve node-visit
frequencies while randomizing the temporal sequence of arrivals yield nearly identical
avalanche macrostatistics for $\beta < 1$ across memory strengths $q = 0$--$0.6$,
demonstrating that dissipation and redistribution rules dominate over temporal memory
ordering in the regime we can reliably characterize.  On Barab\'{a}si--Albert networks,
fixed per-neighbor transfer is strongly hub-sensitive, while degree-normalized transfer
suppresses runaways but yields distributions better described by exponentials.  The
central conclusion is therefore regime-based: memory-biased driving localizes stress
injection and shapes visitation hotspots, but broad cascade behavior is governed
primarily by stress balance, dissipation strength, and network topology.
\end{abstract}

\maketitle

% -----------------------------------------------------------------------
\section{Introduction}
% -----------------------------------------------------------------------

Random walks on networks provide a natural framework for modeling transport, search,
mobility, and information flow in complex systems~\cite{barrat2008dynamical,%
barabasi2016network,newman2018networks}.  Empirical trajectories, however, are rarely
memoryless.  Returns to previously visited locations are well documented in animal
movement, human mobility, and information spreading
processes~\cite{boyer2014random,rosvall2014memory,kim2016network}.  Preferential-return
walks capture this effect by allowing the walker to relocate to a previously visited node
with probability proportional to accumulated visitation
counts~\cite{evans2011diffusion,boyer2014random,guerrero2025random}.

Coupling such a memory-biased drive to local threshold dynamics yields a tractable model
for stress accumulation and collective relaxation on networks.  Threshold-triggered
redistribution is the defining mechanism of sandpile models, which were originally
introduced as archetypal examples of self-organized criticality
(SOC)~\cite{bak1987self,jensen1998self}.  On networks, sandpile phenomenology is known to
depend strongly on topology and on the precise formulation of redistribution rules,
particularly for scale-free and small-world
graphs~\cite{goh2003sandpile,lahtinen2005sandpiles}.  The broad avalanche distributions
that arise in such settings naturally invite SOC language, but modern
analysis~\cite{dickman1998soc,vespignani1998how,clauset2009power,stumpf2012critical,%
markovic2014power} makes clear that broad tails alone do not establish asymptotic
criticality: conservation, dissipation, finite-size scaling, and topology must each be
assessed carefully.

The present study adopts a correspondingly narrow scope.  We ask which components of the
coupled model are responsible for broad cascade distributions, and we subject candidate
explanations to explicit control tests.  The answer that emerges is not that memory
generates SOC.  Rather, fixed transfer rules are brittle on homogeneous networks, small
dissipation stabilizes a broad but subcritical finite-cascade regime on Watts--Strogatz
(WS) graphs, and hub-sensitive redistribution rules can produce misleading results on
heterogeneous topologies.  Furthermore, once node-visit marginals are held fixed,
randomizing the temporal order of visits alters avalanche macrostatistics only weakly in
the dissipative regime, substantially reducing the scope of any memory-centered
interpretation.

% -----------------------------------------------------------------------
\section{Model}
% -----------------------------------------------------------------------

We consider a connected undirected graph $G = (V, E)$ with $N$ nodes.  At each
discrete time step $t$ a single walker occupies node $i_t$.  Each node $i$ carries a
cumulative visitation count $v_i(t)$ and a stress variable $s_i(t)$.

% ---
\subsection{Memory-Biased Drive}
% ---

At each step the walker moves according to a two-mode rule:
\begin{itemize}
  \item with probability $1 - q$, it steps to a uniformly chosen neighbor of its current
        node (unbiased diffusion);
  \item with probability $q$, it resets to a previously visited node $k$ drawn from the
        distribution
        \begin{equation}
          P_{\mathrm{reset}}(k,t) = \frac{v_k(t)}{\displaystyle\sum_{\ell \in V} v_\ell(t)}.
          \label{eq:reset}
        \end{equation}
\end{itemize}
After each move the visitation count of the arrival node is incremented.  The parameter
$q \in [0,1]$ controls the strength of preferential memory: $q = 0$ recovers an ordinary
random walk, while $q \to 1$ concentrates returns on the most frequently visited nodes.

% ---
\subsection{Stress Dynamics and Redistribution Rules}
% ---

Each arrival adds one unit of stress to the visited node:
\begin{equation}
  s_i(t+1) = s_i(t) + \delta_{i,\,i_{t+1}}.
\end{equation}
If $s_j > T_j$ for any node $j$, that node \emph{topples}: its stress is redistributed
to neighbors, possibly triggering further topplings.  We study three redistribution
rules.

\paragraph{Fixed per-neighbor transfer.}
A toppling node releases its entire stress and deposits a fixed amount $\alpha$ on each
neighbor:
\begin{align}
  s_j &\leftarrow 0, \label{eq:fixed_self}\\
  s_\ell &\leftarrow s_\ell + \alpha, \qquad \forall\,\ell \in \mathcal{N}(j).
  \label{eq:fixed_nbr}
\end{align}

\paragraph{Degree-normalized transfer.}
Each neighbor receives a fraction of the redistributed stress determined by the degree
$k_j$ of the toppling node:
\begin{equation}
  s_\ell \leftarrow s_\ell + \frac{\alpha}{k_j},
  \label{eq:normalized}
\end{equation}
so that $\alpha$ represents the total stress released rather than a per-neighbor
increment.

\paragraph{Subtractive dissipative transfer.}
The toppling node loses exactly $T_j$ units, and only a fraction $\beta$ of that amount
is redistributed:
\begin{align}
  s_j &\leftarrow s_j - T_j, \label{eq:dissipative_self}\\
  s_\ell &\leftarrow s_\ell + \frac{\beta T_j}{k_j},
  \qquad \forall\,\ell \in \mathcal{N}(j), \label{eq:dissipative_nbr}
\end{align}
with $0 < \beta \leq 1$.  The case $\beta = 1$ is locally conservative, while $\beta < 1$
introduces controlled dissipation.

The \emph{avalanche size} $S$ is the total number of topplings triggered by a single
walker move.  All topplings are resolved deterministically using a first-in, first-out
queue until no node satisfies $s_i > T_i$.

% ---
\subsection{Simulation Protocol}
% ---

We simulate Watts--Strogatz (WS) graphs with mean degree $k = 4$ and rewiring
probability $p$~\cite{watts1998collective}, and Barab\'{a}si--Albert (BA) graphs with
attachment parameter $m = 2$~\cite{barabasi1999emergence}.  Thresholds are uniform,
$T_i = T = 6$, unless stated otherwise.  Drive steps with $S = 0$ are retained when
computing event rates but excluded from tail fits.

The primary diagnostics are:
\begin{itemize}
  \item the fraction of non-zero avalanche events;
  \item the fraction of system-scale events, defined as $S \geq 0.1 N$;
  \item tail fits by discrete maximum-likelihood estimation with AIC-based comparison
        against exponential and lognormal
        alternatives~\cite{clauset2009power,alstott2014powerlaw};
  \item a bootstrap Kolmogorov--Smirnov (KS) test of the fitted power-law
        hypothesis~\cite{clauset2009power};
  \item extended observables in the subtractive regime: avalanche duration, affected
        area, and a branching-ratio proxy defined as the mean number of newly unstable
        nodes created per toppling event.
\end{itemize}

To isolate the contribution of temporal memory ordering, we also construct
\emph{shuffled-order controls}: the original arrival sequence is randomly permuted while
preserving the empirical node-visit frequencies.  Comparing original and permuted dynamics
separates the effect of memory-driven visit frequencies from the effect of the specific
temporal order in which those visits occur.

% -----------------------------------------------------------------------
\section{Results}
% -----------------------------------------------------------------------

% ---
\subsection{Fixed Transfer on WS Graphs Is Broad but Brittle}
% ---

For WS networks with $k = 4$ and $T = 6$, the stress budget of the fixed per-neighbor
rule is governed by the balance between the stress removed at a toppling ($T$) and the
total stress injected into the neighborhood ($\alpha k$).  A crossover is therefore
expected near $\alpha k \simeq T$, i.e., $\alpha \simeq 1.5$.

\begin{figure}
  \centering
  \includegraphics[width=\linewidth]{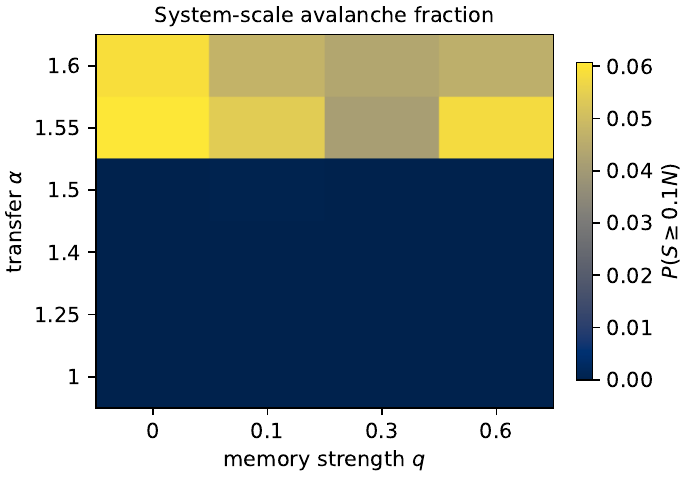}
  \caption{WS regime scan over memory strength $q$ and fixed transfer coefficient
    $\alpha$, with $T = 6$ and $k = 4$.  Color encodes the fraction of drive steps with
    $S \geq 0.1 N$.  Broad cascades emerge near the stress-balance condition
    $\alpha k \simeq T$, with negligible dependence on $q$.}
  \label{fig:phase_diagram}
\end{figure}

\begin{figure}
  \centering
  \includegraphics[width=\linewidth]{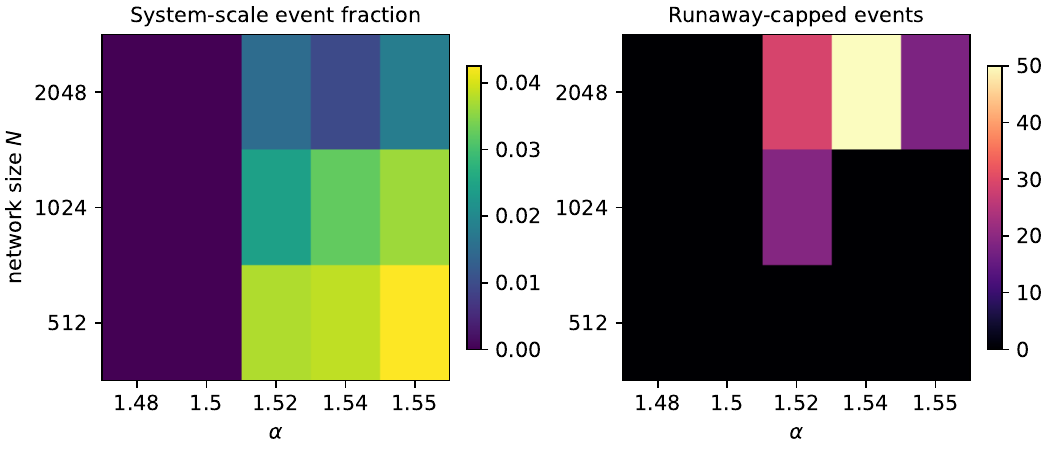}
  \caption{Focused scan of the fixed-transfer transition at $q = 0.3$, $p = 0.1$,
    $T = 6$, and $k = 4$.  Below the transition the system is subcritical; above it the
    fixed rule becomes progressively more unstable with increasing $N$.}
  \label{fig:ws_alpha_window}
\end{figure}

The regime scan in Fig.~\ref{fig:phase_diagram} confirms this picture: system-scale
events are essentially absent for $\alpha \leq 1.5$ and appear sharply above that
boundary, with little dependence on the memory parameter $q$.  The focused scan of
Fig.~\ref{fig:ws_alpha_window} reveals, however, that the broad-cascade window is narrow
and unstable.  Values $\alpha = 1.48$ and $1.50$ are clearly subcritical across $N =
512$--$2048$, whereas $\alpha \geq 1.52$ produces increasingly runaway-capped events as
$N$ grows.  The fixed rule therefore does not support a clean finite-size scaling window
of broad but stable cascades.

% ---
\subsection{Small Dissipation Stabilizes Broad Finite Cascades}
% ---

\begin{figure}
  \centering
  \includegraphics[width=\linewidth]{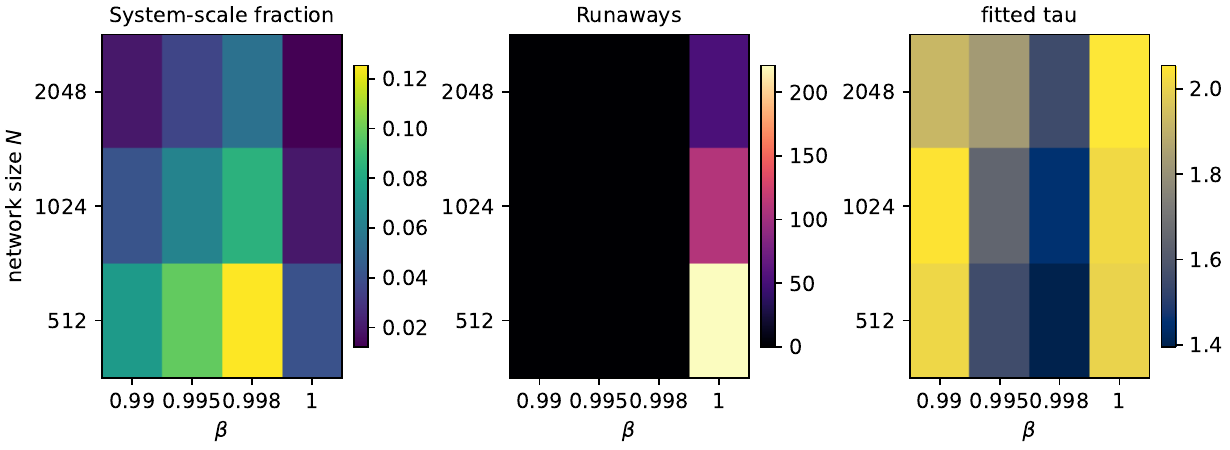}
  \caption{WS results for the subtractive dissipative rule.  The dissipative window
    $\beta = 0.995$--$0.998$ remains broad and non-runaway through $N = 2048$, while the
    conservative limit $\beta = 1$ becomes runaway-capped at these sizes.}
  \label{fig:adaptive_subtractive}
\end{figure}

The subtractive dissipative rule changes the picture qualitatively.
Figure~\ref{fig:adaptive_subtractive} shows that $\beta = 0.995$ and $0.998$ sustain
broad, non-runaway cascades through $N = 2048$, whereas the conservative limit $\beta =
1$ develops runaway-capped events at the same sizes.  This constitutes the strongest
affirmative result of the study: even a very small per-toppling dissipation ($0.2$--$0.5\%$)
is sufficient to stabilize large finite cascades that the fixed conservative rule cannot
sustain.

\begin{figure}
  \centering
  \includegraphics[width=\linewidth]{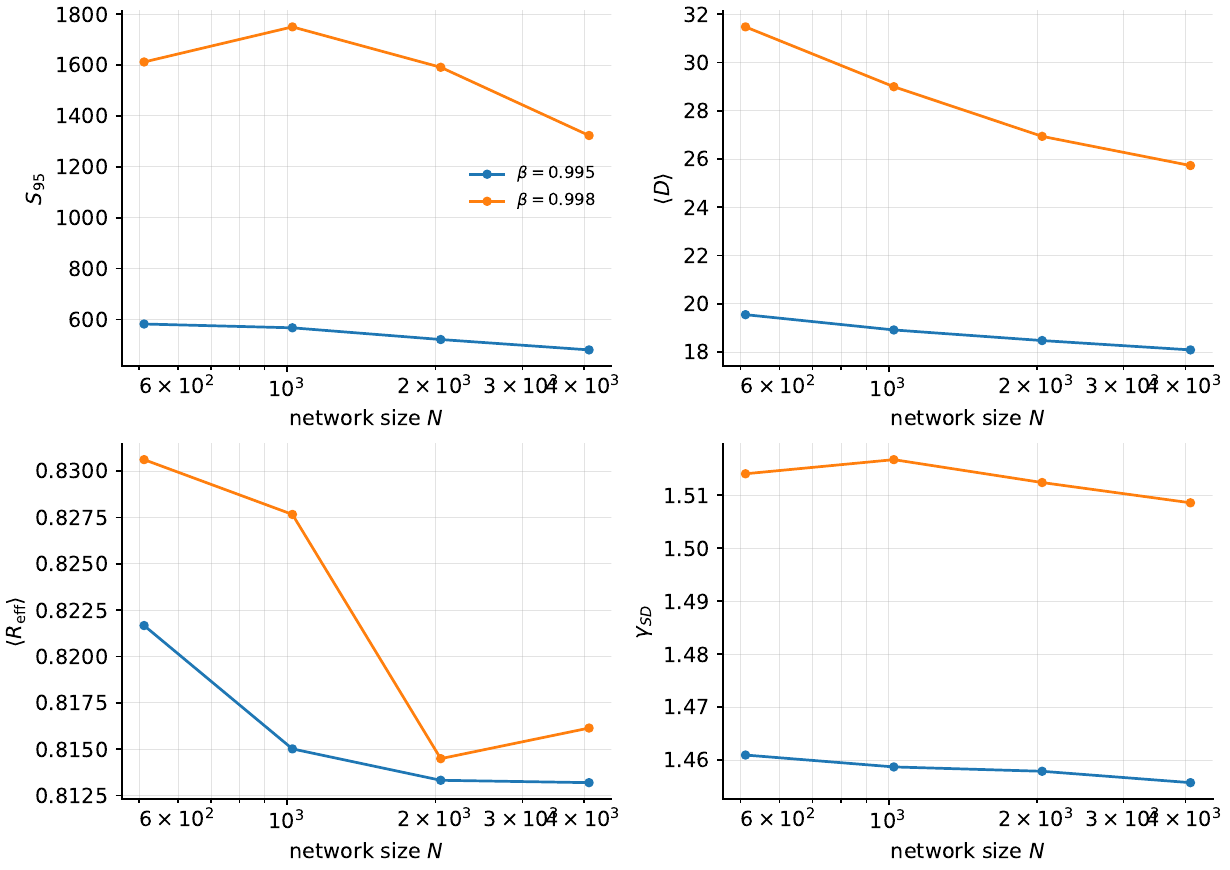}
  \caption{Extended finite-size validation of the subtractive rule through $N = 4096$.
    Both dissipative branches remain non-runaway and power-law-preferred by AIC, but
    system-scale event fractions decrease with $N$ and the branching proxy remains below
    one throughout.}
  \label{fig:phase_a_scaling}
\end{figure}

Extended validation at $N = 4096$ (Fig.~\ref{fig:phase_a_scaling}) clarifies the
asymptotic behavior.  Both $\beta = 0.995$ and $0.998$ remain non-runaway and continue
to favor power-law tails by AIC.  However, the fraction of system-scale events decreases
monotonically with $N$, mean avalanche duration is approximately size-independent, the
branching-ratio proxy stabilizes around $0.81$--$0.83$ (well below the critical threshold
of unity), and the 95th-percentile avalanche size shows no systematic growth with $N$.
Combined with $p_{\mathrm{boot}} < 0.01$ at every tested system size, these results are
inconsistent with asymptotic criticality and instead support the interpretation of a
\emph{broad finite dissipative regime}: avalanche distributions are heavy-tailed at
accessible scales but do not converge to a critical fixed point as $N \to \infty$.

% ---
\subsection{Temporal Memory Ordering Has a Weak Macroscopic Effect}
% ---

A natural question is whether the broad dissipative cascade regime is driven specifically
by the temporal ordering of arrivals produced by the memory walk, or whether it arises
primarily from the visit-frequency distribution that memory induces.  The shuffled-order
controls provide a direct answer.

\begin{figure}
  \centering
  \includegraphics[width=\linewidth]{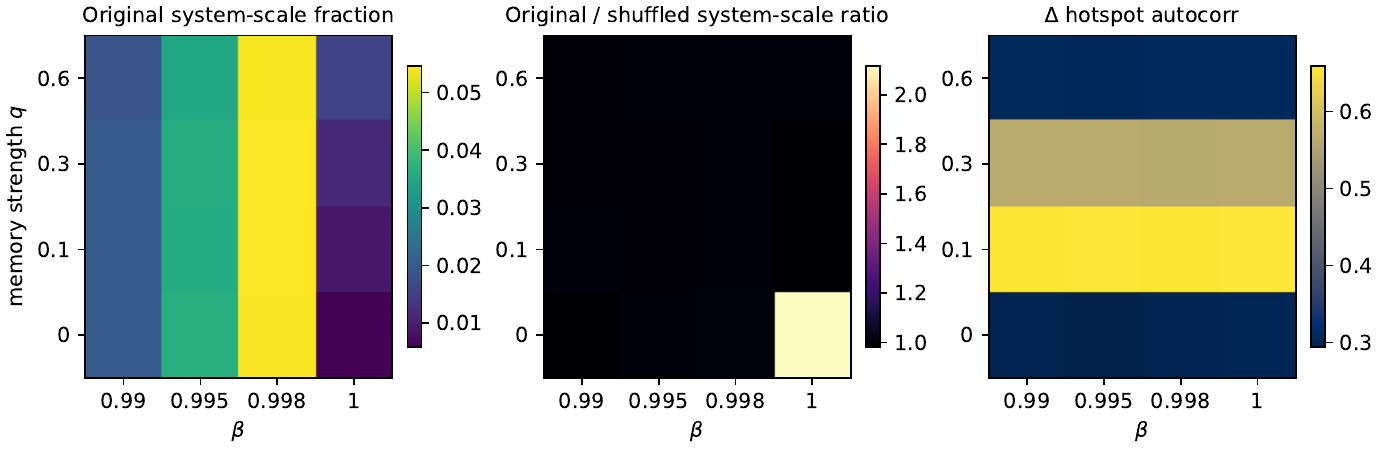}
  \caption{Shuffled-order control for the subtractive WS model at $N = 2048$.
    For $\beta < 1$, preserving node-visit marginals while randomizing arrival order
    leaves all macroscopic avalanche observables nearly unchanged across $q = 0$--$0.6$.}
  \label{fig:phase_b_heatmaps}
\end{figure}

Figure~\ref{fig:phase_b_heatmaps} compares macroscopic avalanche observables for the
original memory-biased sequence against randomly permuted sequences with identical
node-visit marginals.  For $\beta = 0.99$, $0.995$, and $0.998$, the system-scale event
fraction changes by less than $2\%$ across all tested values of $q = 0$, $0.1$, $0.3$,
and $0.6$; the branching-ratio proxy is similarly insensitive to the permutation.  Only
in the conservative case $\beta = 1$ does temporal ordering produce a visible effect,
and even then every value of $q$ still generates runaway-capped cascades.

These results indicate that, within the dissipative regime, memory biases the
\emph{spatial} distribution of stress injection---concentrating load on frequently
revisited nodes and sustaining visitation hotspots---but the specific temporal ordering
of those injections plays only a minor role in determining avalanche macrostatistics.

% ---
\subsection{BA Networks: Hub Sensitivity and Transfer Rule Design}
% ---

Heterogeneous degree distributions introduce additional complications.  Under the fixed
per-neighbor rule on BA networks, a hub toppling injects total stress proportional to its
degree, making the rule strongly hub-sensitive and capable of driving supercritical
behavior through a small number of high-degree nodes.  Degree-normalized transfer removes
this artifact by bounding the total stress released per toppling event at $\alpha$,
regardless of degree.

\begin{figure}
  \centering
  \includegraphics[width=\linewidth]{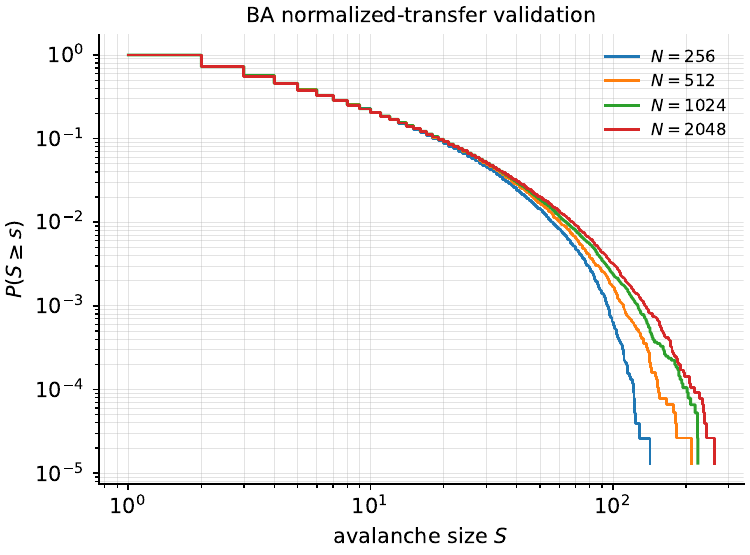}
  \caption{Long-run BA validation with degree-normalized transfer across $N =
    256$--$2048$.  Runaways are suppressed, but tail fits favor exponentials over power
    laws.}
  \label{fig:ba_network}
\end{figure}

The long-run validation with degree-normalized transfer
(Fig.~\ref{fig:ba_network}) confirms that runaways are effectively suppressed across $N =
256$--$2048$, with stable non-zero avalanche rates throughout.  Discrete maximum-likelihood
analysis, however, favors exponential over power-law tails in the validation run.  The
normalized rule therefore serves as a necessary stabilization, but not as evidence for
scale-free avalanche behavior on BA graphs.

% ---
\subsection{Summary of Tail Diagnostics}
% ---

Table~\ref{tab:diagnostics} collects key validation statistics for the main cases
discussed above.  In all cases the bootstrap KS test rejects a pure power law
($p_{\mathrm{boot}} < 0.01$), while AIC favors power-law tails for the WS subtractive
cases and exponential tails for the BA case.  Together with the finite-size behavior
described above, these results support the broad finite dissipative characterization for
WS and argue against scale-free BA criticality.

\begin{table}
\caption{Validation diagnostics for the main simulation cases. The preferred tail model
  is selected by AIC from power law, exponential, and lognormal candidates.
  All cases yield $p_{\mathrm{boot}}<0.01$ from the bootstrap KS test.
  WS\,sub refers to the subtractive dissipative rule on Watts--Strogatz graphs;
  BA\,norm refers to degree-normalized transfer on Barab\'{a}si--Albert graphs.}
\label{tab:diagnostics}
\begin{ruledtabular}
\begin{tabular}{cccccc}
Case & $\beta$ & $N$ & $\hat{\tau}$ & KS & AIC \\
\hline
WS fixed  & ---   & 2048 & $1.513\pm0.003$ & 0.068 & power law \\
WS sub    & 0.995 & 2048 & $1.736\pm0.005$ & 0.070 & power law \\
WS sub    & 0.995 & 4096 & $1.865\pm0.005$ & 0.067 & power law \\
WS sub    & 0.998 & 2048 & $1.513\pm0.003$ & 0.062 & power law \\
WS sub    & 0.998 & 4096 & $1.538\pm0.003$ & 0.055 & power law \\
BA norm   & ---   & 2048 & $2.300\pm0.010$ & 0.127 & exp.      \\
\end{tabular}
\end{ruledtabular}
\end{table}

% -----------------------------------------------------------------------
\section{Discussion}
% -----------------------------------------------------------------------

The results collectively support a more precise and conservative interpretation than
simple SOC language would suggest.  Broad cascades do occur across a range of parameters,
but the dominant controls are dissipation strength, stress balance, and topology---not the
memory parameter $q$.  The fixed transfer rule is brittle on WS graphs and hub-sensitive
on BA graphs; neither pathology is resolved by increasing $q$.  Small dissipation in the
subtractive rule, by contrast, stabilizes a broad finite cascade regime on WS networks
that persists to the largest system sizes tested while avoiding runaways.

This phenomenology is more naturally described in terms of dissipative or quasi-critical
network dynamics than in terms of a conservative SOC fixed
point~\cite{bonachela2010self,kinouchi2019stochastic}.  The clearest positive result of
the study is the identification of a redistribution rule---the subtractive dissipative
rule with $\beta \lesssim 1$---that achieves broad cascade distributions while suppressing
the runaway instability that limits the fixed rule.

The shuffled-order control results further refine the interpretation.  Memory-biased
walking does influence the cascade statistics, but primarily through its effect on
node-visit frequencies rather than through the specific temporal sequencing of arrivals.
Increasing $q$ concentrates stress accumulation on frequently revisited nodes, shaping
hotspot persistence and potentially the spatial structure of cascades; however, these
structural effects are already encoded in the marginal visit distribution, and the
additional signal carried by temporal ordering is small in the dissipative regime.  A
memory-centered explanation of broad cascades should therefore be qualified: memory
matters as a structured drive that modifies the effective stress landscape, but the
temporal signature of memory does not appear to be the primary mechanism.

These distinctions are practically relevant in settings where non-conservative loading
and heterogeneous structure are unavoidable, including neural
networks~\cite{beggs2003neuronal} and infrastructure
systems~\cite{barrat2008dynamical}.  A model that cleanly separates the contributions
of drive structure, redistribution design, and topology provides a useful reference for
interpreting cascade phenomena in those contexts.

% -----------------------------------------------------------------------
\section{Conclusion}
% -----------------------------------------------------------------------

We have studied a coupled model of memory-biased random-walk driving and sandpile-like
stress relaxation on Watts--Strogatz and Barab\'{a}si--Albert networks.  The principal
findings are as follows.

The fixed per-neighbor transfer rule on WS graphs produces broad cascades only in a
narrow parameter window near the stress-balance condition $\alpha k \simeq T$.  That
window is not stable: slightly subcritical parameters yield no broad cascades, while
slightly supercritical parameters generate system-spanning runaways at large $N$.

The subtractive dissipative rule with $\beta = 0.995$--$0.998$ stabilizes broad,
non-runaway cascades through $N = 4096$.  Extended finite-size diagnostics---decreasing
system-scale event fractions, size-independent durations, branching ratios below one, and
bootstrap rejection of pure power laws---place this behavior in a broad finite
dissipative regime rather than at a true SOC critical point.

Shuffled-order controls establish that, once node-visit marginals are held fixed, temporal
memory ordering has only a weak influence on avalanche macrostatistics for $\beta < 1$.
Memory-biased driving shapes the spatial distribution of stress injection, but broad
cascade behavior is governed primarily by the redistribution rule and network topology.

On BA networks, fixed transfer is hub-sensitive and must be replaced by a
degree-normalized rule to avoid artificial supercriticality; the resulting distributions
are better characterized as exponential than power-law.

Taken together, these results suggest that this class of models is best understood as a
framework for exploring \emph{network cascade regimes} rather than as a candidate
mechanism for self-organized criticality.  The most useful contribution is the
identification of a dissipative redistribution design that achieves robust broad cascades
without runaway instability---a property likely to be relevant beyond the specific network
models studied here.

\bibliographystyle{apsrev4-2}
\bibliography{references}

\end{document}